\documentclass[prd,aps,preprint,tightenlines,superscriptaddress,nofootinbib,showpacs,showkeys,floatfix]{revtex4-1}
\usepackage{amsfonts}
\usepackage{array}
\usepackage{epsfig}
\usepackage{rotating}
\usepackage{multirow}
\usepackage{soul}
\usepackage{amsmath}    
\usepackage{verbatim}   
\usepackage{color}      
\usepackage{colordvi}
\usepackage[dvipsnames]{xcolor}
\usepackage[colorlinks = true,
            linkcolor = blue,
            urlcolor  = blue,
            citecolor = blue,
            anchorcolor = blue]{hyperref}

\usepackage{graphicx}

\usepackage{url}

\def\muf{{\mu^{}_f}}
\def\mufs{{\mu^{\,2}_f}}
\def\mur{{\mu^{}_r}}

\newcommand{\gsim}{\;\rlap{\lower 3.5 pt \hbox{$\mathchar \sim$}} \raise 1pt
 \hbox {$>$}\;}
\newcommand{\lsim}{\;\rlap{\lower 3.5 pt \hbox{$\mathchar \sim$}} \raise 1pt
 \hbox {$<$}\;}

\graphicspath{{./figs/}}

\usepackage{pslatex}
\usepackage{txfonts}
\usepackage[latin1]{inputenc}
\usepackage[T1]{fontenc}

\begin{document}
\begin{titlepage}
\thispagestyle{empty}
\noindent
CERN-TH-2024-222
\hfill
December 2024 \\
\noindent
DESY 24-207\\
\noindent
P3H-24-104 \\
\noindent
TTP24-048
\vspace{3.0cm}

\begin{center}
  {\bf \Large
  Updated predictions for toponium production at the LHC 
  }
  \vspace{1.25cm}

 {\large
   M. V. Garzelli$^{\, a,b}$,
   G. Limatola$^{\, b,c}$,
   S.-O.~Moch$^{\, b}$,
   M. Steinhauser$^{\, d}$,
   O. Zenaiev$^{\, b}$
 }
 \\
 \vspace{1.25cm}
 {\it
$^a$ CERN, Department of Theoretical Physics, 
CH-1211 Geneva 23, Switzerland\\[1ex]
$^b$ Universit\"at Hamburg, II. Institut f\"ur Theoretische Physik, \\
Luruper Chaussee 149, 22761 Hamburg, Germany \\[1ex]
$^c$ Deutsches Elektronen-Synchrotron DESY
Notkestra{\ss}e 85, 22607 Hamburg, Germany \\[1ex]
$^d$ Karlsruhe Institute of Technology (KIT),
Institut f\"ur Theoretische Teilchenphysik,
Wolfgang-Gaede Stra{\ss}e 1, 76131 Karlsruhe, Germany\\[1ex]
 }

\vspace{1.4cm}
\large {\bf Abstract}
\vspace{-0.2cm}
\end{center}
We provide an update on QCD predictions for top-quark pair production close to threshold including bound state effects at the Large Hadron Collider. 
We compute the top-quark pair invariant mass distribution $d\sigma/dM_{t\bar{t}}$, including Coulomb resummation for bound-state effects, as well as threshold resummation for emissions of soft and collinear gluons. 
We discuss uncertainty estimates and present a proposal for the use of these predictions in experimental analyses.

\end{titlepage}

\newpage
\setcounter{footnote}{0}
\setcounter{page}{1}
A recent analysis of top-quark pair-production using the invariant mass $M_{t\bar{t}}$ of the $t\bar{t}$-pair and angular variables sensitive to its spin, 
by the CMS collaboration at the Large Hadron Collider (LHC)~\cite{CMS:2024ynj} 
with the full integrated luminosity from Run 2, updating ~\cite{CMS:2019pzc},
has identified a highly significant excess of events at an invariant mass $M_{t\bar{t}}$ of about twice the value of the top-quark pole mass $m_t$.
This intriguing result may be due to the production of a new heavy beyond-the-Standard-Model (BSM) Higgs boson decaying into a  $t\bar{t}$ pair. 
Such a search has also been conducted by the ATLAS collaboration, but with less sensitivity to the region $M_{t\bar{t}} \simeq 2 m_t$, so that only exclusion limits for various BSM scenarios have been reported~\cite{ATLAS:2024vxm}.

It is well known that within the Standard Model (SM), an enhanced cross section $d\sigma/dM_{t\bar{t}}$ is expected due to bound-state effects in quantum chromodynamics (QCD), arising from gluon exchanges in $t\bar{t}$ pairs near  threshold~\cite{Bigi:1986jk,Fadin:1987wz,Fadin:1990wx}. 
Therefore, accurate theoretical predictions within the SM are crucial to support reliable searches for BSM physics. 
Several dedicated studies of the bound-state effects, which are not included in fixed-order QCD predictions or in the standard parton shower Monte Carlo approaches used in experimental analyses, have been conducted in the past~\cite{Hagiwara:2008df,Kiyo:2008bv,Sumino:2010bv}.

In this letter, we provide an update of the QCD predictions for $t\bar{t}+X$ production at the LHC, previously reported in Ref.~\cite{Kiyo:2008bv}, in order to match the Run 2 kinematics and the settings of the CMS analysis~\cite{CMS:2024ynj}.
In particular, we focus on predictions for $d\sigma/dM_{t\bar{t}}$ around threshold, including bound state effects due to the exchange of Coulomb-like gluons between the top and the anti-top quark, as well as the real emission of soft and collinear gluons. 
Other recent approaches have relied on effective field theory methods to obtain these QCD predictions~\cite{Ju:2020otc}. In view of the experimental prospects, phenomenological studies based on models  have also addressed bound-state signatures~\cite{Fuks:2021xje,Maltoni:2024tul} and characterized them through different $t\bar{t}$ decay channels (dileptonic vs. semileptonic)~\cite{Aguilar-Saavedra:2024mnm}, while more recent work~\cite{Fuks:2024yjj} has focused on re-weighting matrix elements for $t\bar{t}$ production and combining them with parton showering in order to assist experimental analyses.

Let us briefly summarize the theory framework of Ref.~\cite{Kiyo:2008bv}, which will be adopted in this letter. 
The $M_{t\bar{t}}$ distribution in $pp$ collisions factorizes into a convolution of a luminosity function ${\cal L}_{ij}$, encoding the long-distance parton dynamics of the initial hadrons, with a short-distance partonic cross section $d{\hat \sigma}_{ij\to T}/dM_{t\bar{t}}$ for the production of a $t\bar{t}$ pair,
\begin{equation}
 M_{t\bar{t}} {{\rm d}\sigma_{P_1 P_2\to T}\over {\rm d}M_{t\bar{t}}}(S, M_{t\bar{t}}^2) 
 =\sum_{i,j} \int_\rho^1 {\rm d}\tau\,
  \bigg[\frac{{\rm d}{\cal L}_{ij}}{{\rm d}\tau}\bigg](\tau,\mufs)
  ~M_{t\bar{t}}\frac{{\rm  d}\hat\sigma_{ij\to T}}{{\rm d}M_{t\bar{t}}}(\hat s,M_{t\bar{t}}^2,\mufs)\,.
 \label{eq:TheConvolution}
\end{equation}
Here, $S$ is the square of the $pp$ center-of-mass energy, $\rho = M_{t\bar{t}}^2/S$, 
$\mu_f$ is the factorization scale, and the `hat' symbol denotes partonic quantities; 
$\tau = \hat{s}/S$, where $\hat{s}$ is the partonic center-of-mass energy squared. 
$T$ denotes a state $^{2S+1}L_J^{[1,8]}$, in spectroscopic notation, with spin $S$, orbital angular momentum $L$, total angular momentum $J$ and color, with the superscript $[1,8]$ denoting a color-singlet or a color-octet configuration. 
The sum is performed over all contributing initial-state parton channels and the luminosity function is given by 
\begin{eqnarray}
  \bigg[\frac{d{\cal L}_{ij}}{d\tau}\bigg](\tau,\mufs)
  &=&
  \int_0^1 {\rm d}x_1 \int_0^1 {\rm d}x_2\, 
  f_{i/P_1}(x_1,\mufs) f_{j/P_2}(x_2, \mufs)\,
  \delta(\tau - x_1 x_2)
  \,,
  \label{eq:lumi}
\end{eqnarray}
in terms of the standard parton distributions (PDFs) $f_{i,j}$ of the two colliding initial-state partons $i$ and $j$, with longitudinal momentum fractions $x_i$ and $x_j$, respectively.

Bound-state effects can be described in non-relativistic QCD (NRQCD), which is valid for small top-quark velocities $\beta_t = \sqrt{1-4m_t^2/M_{t\bar{t}}^2}$ in the $t\bar{t}$ rest frame.
In NRQCD the partonic cross section $d{\hat \sigma}_{ij\to T}/dM_{t\bar{t}}$ can be factorized into a product of a hard function $F_{ij \to T}$ multiplied by the imaginary part of non-relativistic Green's functions
$G^{[1,8]}(M_{t\bar{t}}+i\Gamma_t) \equiv G^{[1,8]}(\vec r = 0, M_{t\bar{t}}+i\Gamma_t)$, evaluated here at zero distance, 
\begin{eqnarray}
  M_{t\bar{t}}\frac{{\rm d}\hat\sigma_{ij\to T}}{{\rm d}M_{t\bar{t}}}(\hat s,M_{t\bar{t}}^2,\mufs)
 &=&
 F_{ij\to T}(\hat{s},M_{t\bar{t}}^2,\mufs)\,
\, \frac{1}{m_t^2}\,{\rm Im} \,G^{[1,8]}(M_{t\bar{t}}+i\Gamma_t)\, .
\label{eq:PartonicXS}
\end{eqnarray}
%
%
The non-relativistic Green's functions $G^{[1,8]}$, considered here for $S$-waves ($L=0$), 
depend on the top-quark width $\Gamma_t$ and on the color state of the $t\bar{t}$ pair. 
They are obtained from the solution of a Schr{\"o}dinger equation accounting for the exchange of potential gluons among the top and anti-top quark. 
The color-singlet Green's function $G^{[1]}$ feels an attractive force, so that the $t\bar{t}$ pair transition into a quasi-bound state is favored. 
This state is colloquially referred to as `toponium', although it should be noted that the top quark decays much faster than it can hadronize, i.e.  $\Gamma_t \gg \Lambda_{\rm{QCD}}$, and thus a proper bound state cannot form, unlike the more common charmonia and bottomonia.
The color-octet Green's function $G^{[8]}$, on the other hand, is governed by repulsion and the $t\bar{t}$ pair does not develop a bound state.
We apply the QCD potential to next-to-leading order (NLO) accuracy~\cite{Fischler:1977yf,Billoire:1979ih}.

The hard function  $F_{ij \to T}$ can be computed in perturbative QCD and 
analytical expressions at NLO can be inferred from Ref.~\cite{Petrelli:1997ge}. 
At NLO accuracy the Green's functions $G^{[1,8]}$ in Eq.~(\ref{eq:PartonicXS}) and the convolution ${\cal L} \otimes F$ of the hard function $F_{ij \to T}$ with the luminosity are individually independent of the renormalization scale $\mu_r$.
Beyond NLO, the factorization formula in Eq.~(\ref{eq:PartonicXS}) needs generalizations, also in case of other differential distributions.
Following Ref.~\cite{Kiyo:2008bv}, our strategy for the computation of the `toponium' cross sections according to Eq.~(\ref{eq:TheConvolution}) combines predictions for the convolution ${\cal L} \otimes F$ with the NRQCD solutions for the Green's functions $G^{[1,8]}$ in the threshold region.
For larger invariant masses $M_{t\bar{t}}$, this approach is not valid and 
predictions in fixed-order perturbation theory, available at next-to-next-to-leading order (NNLO), can be directly applied. 
To that end, we rely on recent work~\cite{Garzelli:2023rvx}, interfacing 
\texttt{MATRIX}~\cite{Catani:2019hip, Grazzini:2017mhc} to \texttt{PineAPPL}~\cite{Carrazza:2020gss}.
The transition between the regions where either NRQCD or fixed-order perturbation theory is applicable requires a matching prescription, obviously.

The hard function  $F_{ij \rightarrow T}$ contains threshold logarithms due to emissions of soft and collinear gluons.
They become large close to the partonic threshold $z = M_{t\bar{t}}^2/\hat{s} \rightarrow 1$ 
and dominate the convolution ${\cal L} \otimes F$ with the luminosity function in Eq.~(\ref{eq:TheConvolution}) near the endpoint $\tau = \rho$, which implies the limit $z \rightarrow 1$.
We resum these logarithms up to next-to-leading logarithmic (NLL) accuracy for the three most relevant contributions to the cross section, corresponding to the channels $gg \rightarrow {^1}S_0^{[1]},\, {^1}S_0^{[8]}$ and $q\bar{q}\rightarrow {^3}S_1^{[8]}$, and accounting for the color-singlet and octet structures of the final-state quark-anti quark system. 
Resummation is conveniently performed in Mellin (i.e. $N$) space, since the cross section factorizes into a product of functions in the relevant (soft, collinear, etc.) phase-space for multiple emissions. 
The functions $F_{ij \rightarrow T}^N$ in $N$-space are obtained by a Mellin transform, calculating the $N$-th Mellin moment with respect to $z$, 
\begin{equation}
\label{eq:MellinN-Fres}
F^N_{ij\to T}(M_{t\bar{t}}^2,\mufs) =  \int\limits_{0}^{1}\,dz\, z^{N-1}\, F_{ij\to T}(\hat s, M_{t\bar{t}}^2,\mufs)
\, .
\end{equation}
The predictions for $d\sigma/dM_{t\bar{t}}$ are then recovered by an inverse Mellin transform of the convolution $({\cal L} \otimes F)^N = {\cal L}^N \cdot F^N$ with the resummed formulae for $F^N_{ij\to T}$ and the Mellin transform ${\cal L}^N$ of the luminosity function. 
This includes matching of the resummed results to the fixed-order ones at NLO, abbreviated as `NLO+NLL', 
according to
%
%
\begin{eqnarray}
\label{eq:defsigmares}
F^{\rm NLO+NLL}_{ij\to T}(\hat s,M_{t\bar{t}}^2,\mufs) &=& 
\int\limits_{c-{\rm i}\infty}^{c+{\rm i}\infty}\, {dN \over 2\pi i}\, x^{-N}\,
\left( F^N_{ij \to T}(M_{t\bar{t}}^2,\mufs) - \left. F^N_{ij \to T}(M_{t\bar{t}}^2,\mufs) \right|_{\rm NLO} \right)
\nonumber\\ &&
  + F^{\rm NLO}_{ij \to T}(\hat s,M_{t\bar{t}}^2,\mufs)\, .
\end{eqnarray}
In this matching, $F^{\rm NLO}_{ij \to T}$ denotes the standard NLO fixed order cross section, 
whereas $F^N_{ij \to T}\bigr|_{\rm NLO}$ is the perturbative truncation of Eq.~(\ref{eq:MellinN-Fres}) at the same order in $\alpha_s$.
Eq.~(\ref{eq:defsigmares}) reproduces the fixed order results at NLO in QCD and resums soft-gluon effects beyond NLO to NLL accuracy.
Following the set-up of Ref.~\cite{Kiyo:2008bv}, we perform the inverse Mellin transform numerically, using the minimal prescription as well as the matching to NLO accuracy outlined there.

\begin{table}[t]
\begin{center}
\renewcommand{\arraystretch}{1.4}
\begin{tabular}{c|c|c|c|c|c|c}
\hline
&
\multicolumn{3}{|c|}{NLO }&
\multicolumn{3}{c}{resummed}\\ \hline\hline
$gg \,\to\, {^1}S_0^{[1]}$ &
18.2 & 18.7 & 18.3
&
19.4 & 20.5 & 21.1
\\
$gg \,\to\, {^1}S_0^{[8]}$ &
55.8 & 55.2 & 52.8 
& 
60.0 & 61.5 & 62.0
\\
$q\bar q \,\to\, {^3}S_1^{[8]}$ &
21.7 & 22.3 & 22.0
& 
22.4 & 22.4 & 22.0
\\
\hline
\end{tabular}
\caption{\label{tab:nlo-res}
  Comparison of the NLO and NLO+NLL resummed 
  result of the convolution ${\cal L}\otimes F$ (in units $10^{-6}$~GeV$^{-2}$)
  for the LHC configuration $\sqrt{S}=13$~TeV with NNPDF3.1 PDFs
  at the reference point $M_{t\bar{t}}=2m_t$.
  The three columns correspond to the scale choices $\mur=\muf \in \{m_t,\, 2m_t,\, 4m_t\}$. 
}
\renewcommand{\arraystretch}{1.}
\end{center}
\end{table}
\begin{table}[t]
\begin{center}
\renewcommand{\arraystretch}{1.4}
\begin{tabular}{c|c|c|c|c|c|c}
\hline
&
\multicolumn{3}{|c|}{NLO }&
\multicolumn{3}{c}{resummed}\\ \hline\hline
$gg \,\to\, {^1}S_0^{[1]}$ &
19.5 & 19.9 & 19.5
&
20.7 & 21.9 & 22.6
\\
$gg \,\to\, {^1}S_0^{[8]}$ &
59.5 & 58.8 & 56.2 
& 
64.0 & 65.6 & 66.2
\\
$q\bar q \,\to\, {^3}S_1^{[8]}$ &
22.9 & 23.5 & 23.2
& 
23.6 & 23.7 & 23.2
\\
\hline
\end{tabular}
\caption{\label{tab:nlo-res-2}
  Same as Tab.~\ref{tab:nlo-res}, but for the reference point $M_{t\bar{t}}=2m_t - 5$ GeV.
}
\renewcommand{\arraystretch}{1.}
\end{center}
\end{table}
Our updated QCD predictions are provided for Run 2 of the LHC, i.e.\ $pp$ collisions at $\sqrt{S} = 13$~TeV, a top-quark pole mass of $m_t=172.5$~GeV and $\Gamma_t = 1.4$~GeV. As in the CMS analysis~\cite{CMS:2024ynj}, we use the NNPDF3.1 PDF 
set~\cite{NNPDF:2017mvq} at NNLO with the default value of $\alpha_s(M_Z)=0.1180$.
The central renormalization and factorization scales $\mu_r$ and $\mu_f$ are fixed to $\mu_r = \mu_f = 2m_t$. 
In Tables~\ref{tab:nlo-res} and~\ref{tab:nlo-res-2}, we present the relevant values for the convolution ${\cal L} \otimes F$ at NLO and, using resummation, at NLO+NLL accuracy
for two values of $M_{t\bar{t}}$ right at threshold and below: 
Tab.~\ref{tab:nlo-res} for $M_{t\bar{t}}=2m_t$ and Tab.~\ref{tab:nlo-res-2} for $M_{t\bar{t}}=2m_t-5$GeV.
The resummed results are enhanced compared to NLO predictions by up to 10\%, depending on the channel and the scale choice of $\mu_r=\mu_f \in \{m_t,\, 2m_t,\, 4m_t\}$. 
Also the scale stability of the resummed results is somewhat improved.
Other $S$-wave partonic channels contributing at NLO, i.e.\ 
$gg \rightarrow {^3}S_1^{[1,8]}$, $q\bar{q}\rightarrow {^1}S_0^{[1,8]}$, 
$gq \rightarrow {^1}S_0^{[1,8]}$ and $gq \rightarrow {^3}S_1^{[8]}$, 
do not exhibit large threshold logarithms and sum up to a small additional contribution of ${\cal O}(5\%)$, cf.\ Ref.~\cite{Kiyo:2008bv} for details.
In addition, $P$-wave states (i.e. $L=1$) are suppressed by additional powers of the top quark velocity, hence further suppressed at threshold and neglected here.
The  $M_{t\bar{t}}{\rm d}\hat\sigma_{ij\to T}/{\rm d}M_{t\bar{t}}$ cross section values in Eq.~(\ref{eq:PartonicXS}) are obtained 
by multiplying the convolutions ${\cal L} \otimes F$ in Tabs.~\ref{tab:nlo-res} and~\ref{tab:nlo-res-2}
by the value for the corresponding Green's function at its fixed renormalization scale $\mu_s \approx 32\, \mathrm{GeV}$.
The relevant color-singlet and octet values for the settings in Tab.~\ref{tab:nlo-res} are 
\begin{eqnarray}
\label{eq:ImG-over-mt-tab1}
\frac{1}{m_t^2}\, {\rm Im} G^{[1]}_{L=0}(345\mathrm{GeV}+i\Gamma_t) &=& 0.03716 \, ,
\nonumber \\
\frac{1}{m_t^2}\, {\rm Im} G^{[8]}_{L=0}(345\mathrm{GeV}+i\Gamma_t) &=& 0.00373 \, ,
\end{eqnarray}
and for Tab.~\ref{tab:nlo-res-2} 
\begin{eqnarray}
\label{eq:ImG-over-mt-tab2}
\frac{1}{m_t^2}\, {\rm Im} G^{[1]}_{L=0}(340\mathrm{GeV}+i\Gamma_t) &=& 0.01067 \, ,
\nonumber \\
\frac{1}{m_t^2}\, {\rm Im} G^{[8]}_{L=0}(340\mathrm{GeV}+i\Gamma_t) &=& 0.00162 \, .
\end{eqnarray}

In passing let us also address concerns raised in Ref.~\cite{Ju:2020otc} about the application of threshold resummation.
For the kinematics under consideration, threshold logarithms saturate the NLO cross section in the dominant channels to about 70\%, cf.\ Ref.~\cite{Kiyo:2008bv}, with the non-collinear contributions accounting for the remainder. 
In Mellin $N$-space these leading power (LP) logarithms appear as $\alpha_s^n \ln^k N$ with $2n \geq k \geq 1$ and there is a clear hierarchy relative to the next-to-leading power (NLP) corrections $\alpha_s^n (\ln^l N)/N$, where $2n-1 \geq l \geq 1$.
It turns out that the LP and NLP terms provide a lower and upper bound on the exact result.
This is well documented to much higher accuracy, i.e.\ (next-to)$^4$-leading logarithmic (N$^4$LL) accuracy in Refs.~\cite{Das:2019btv,Das:2020adl} for deep-inelastic scattering and the Higgs boson production in gluon-gluon fusion, which is directly 
related to the toponium channel $gg \rightarrow {^1}S_0^{[1]}$.
With the resummation of the LP contributions, the results presented here can be considered as a lower bound on the cross section.
Refinements of the estimates through the systematic treatment of NLP corrections ${\cal O}(1/N)$ are left for future work.\footnote{
Note that the power series (LP, NLP, \dots) in Mellin $N$ space used here and the one in $z = M_{t\bar{t}}^2/\hat{s}$ space (or\ $\beta_t$ space) in Ref.~\cite{Ju:2020otc} based on factorization in Soft-Collinear Effective Theory differ by numerically important sub-leading terms.}

\begin{figure}[t]
\begin{center}
\includegraphics[width=0.7\textwidth]{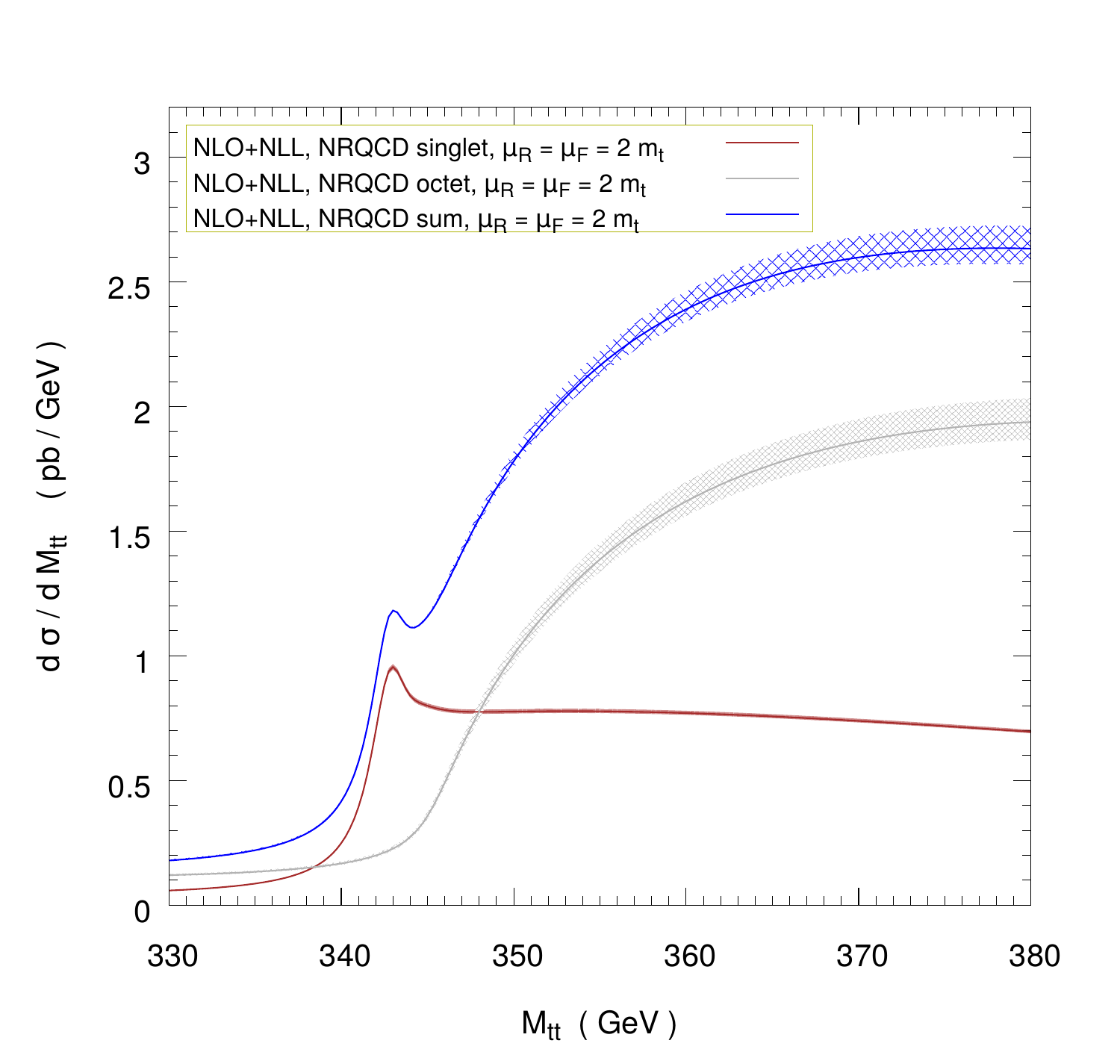}
\caption{
\label{fig:xs18}
Invariant mass distribution $d\sigma/dM_{t\bar{t}}$ in NRQCD with Coulomb and threshold resummation for the dominant 
individual production channels, also listed in Tab.~\ref{tab:nlo-res}:
$gg\rightarrow {^1S_0^{[1]}}$ (brown),  $gg\rightarrow {^1S_0^{[8]}}$ + $q\bar{q}\rightarrow {^3S_1^{[8]}}$ (grey) and their sum (blue).
The width of the band reflects the dependence on the scale choices $\mur = \muf \in \{m_t,\, 2m_t,\, 4m_t\}$ for the convolution ${\cal L}\otimes F$.}
\end{center}
\end{figure}
%
\begin{figure}[h]
  \begin{center}
    \includegraphics[width=0.7\textwidth]{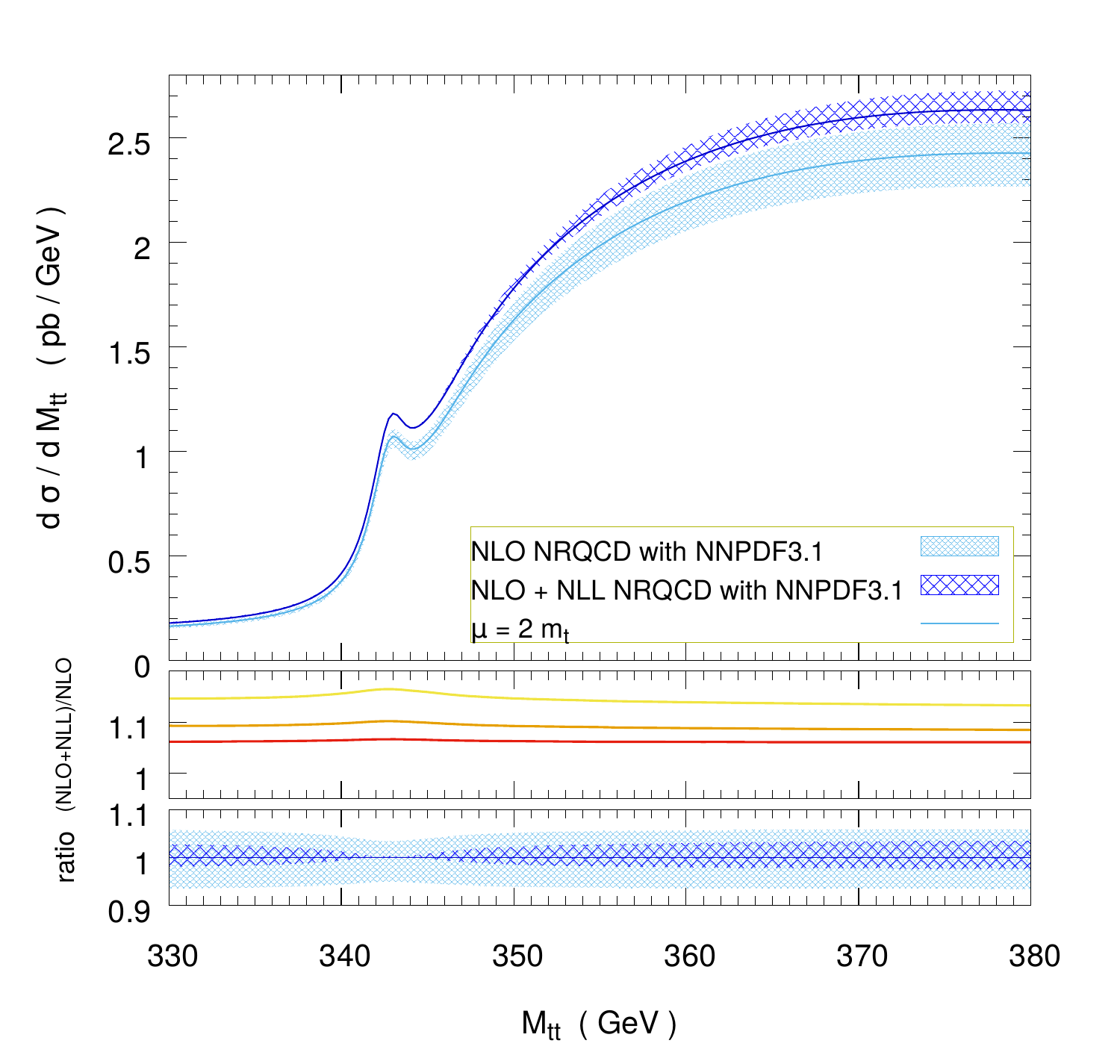}
    \caption{\label{fig:nlo+res}
Same as Fig.~\ref{fig:xs18}, comparing NLO (light-blue) and NLL resummed (blue) predictions for the hard function $F_{ij\to T}$, including scale uncertainties from the range $\mur=\mu_f \in \{m_t,\, 2m_t,\, 4m_t\}$. 
The panel in the middle quantifies the effect of the resummation for predictions at the scales $m_t$ (red), $2 m_t$ (orange), $4 m_t$ (yellow) and the bottom panel shows the ratio of the scale uncertainty bands to the respective central values.
    }
  \end{center}
\end{figure}
%
In a next step, we compute the full partonic and hadronic cross sections in NRQCD according to 
Eqs.~(\ref{eq:TheConvolution}) and (\ref{eq:PartonicXS}).
In Fig.~\ref{fig:xs18} we show the individual color configurations (singlet and octet), combining Coulomb resummation through the Green's functions and threshold resummation at NLL accuracy for the hard function.
The bands illustrate the variation of the convolution ${\cal L}\otimes F$ around the central scale $\mur = \muf = 2m_t$.
While the $gg \rightarrow {^1}S_0^{[1]}$ channel exhibits the resonance associated with the `toponium' bound state, 
the sum of the octet channels $gg \rightarrow {^1}S_0^{[8]}$ and $q\bar{q}\rightarrow {^3}S_1^{[8]}$ show the repulsive effect at threshold for $M_{t\bar{t}} \simeq 2m_t$ and a steeply rising cross section $d\sigma/dM_{t\bar{t}}$, which becomes dominant at 
$M_{t\bar{t}} \gtrsim 350$~GeV. 
The settings for the solution of the Schr{\"o}dinger equation in terms of the Green's functions are the same as in our previous study and we refer to Ref.~\cite{Kiyo:2008bv} for a detailed discussion on the uncertainties, cf. also Refs.~\cite{Beneke:1999qg,Beneke:2015kwa}. 
In particular, we do not vary the soft scale $\mu_s$ in the NRQCD part.
From studies of the related preocess of $e^+e^-$ annihilation, see e.g. Ref.\ \cite{Beneke:2015kwa}, 
it is known that higher-order predictions at NNLO and beyond provide sizeable corrections, well above the range of the NLO soft-scale variations.
Therefore, the variations of $\mu_s$ for the Green's functions $G^{[1,8]}$ underestimate the uncertainty.
From these studies we estimate the color singlet Green's function $G^{[1]}$ to increase by about ${\cal O}(10\%)$ in the peak region due to NNLO corrections.

We also note that the value of the top-quark decay width $\Gamma_t$ has a strong impact on the shape of the invariant mass distribution $d\sigma/dM_{t\bar{t}}$ in Fig.~\ref{fig:xs18} below the threshold $M_{t\bar{t}} \simeq 2m_t$, due to the octet Green's function $G^{[8]}$ falling off slowly for  $M_{t\bar{t}} \lesssim 2m_t$.
Since the underlying NRQCD approach is only valid for small top-quark velocities, its region of validity is restricted to a narrow $M_{t\bar{t}}$ range around threshold, say 
$|M_{t\bar{t}} - 2m_t| \lesssim 5$~GeV. 
Therefore, cross-section predictions based on the NRQCD approach for $d\sigma/dM_{t\bar{t}}$ at values of $M_{t\bar{t}} \simeq 300$~GeV far below threshold are unphysical. 
Hence the conclusions drawn in Ref.~\cite{Ju:2020otc} about an increase in the cross section from this region are invalid.
 
In Fig.~\ref{fig:nlo+res} we compare, for the sum of all contributions (singlet and octet), the 
$M_{t\bar{t}}$ distribution including the impact of the threshold resummation at NLL accuracy compared to fixed-order perturbation theory for the hard function $F_{ij\to T}$. 
The plot illustrates, for the entire $M_{t\bar{t}}$ range, that the resummation increases the cross section by approximately 10\% for the central scale $\mu_r = \mu_f = 2m_t$, and that the scale stability is improved, cf.\ also Tab.~\ref{tab:nlo-res}.

\begin{figure}[h]
\begin{center}
\includegraphics[width=0.7\textwidth]{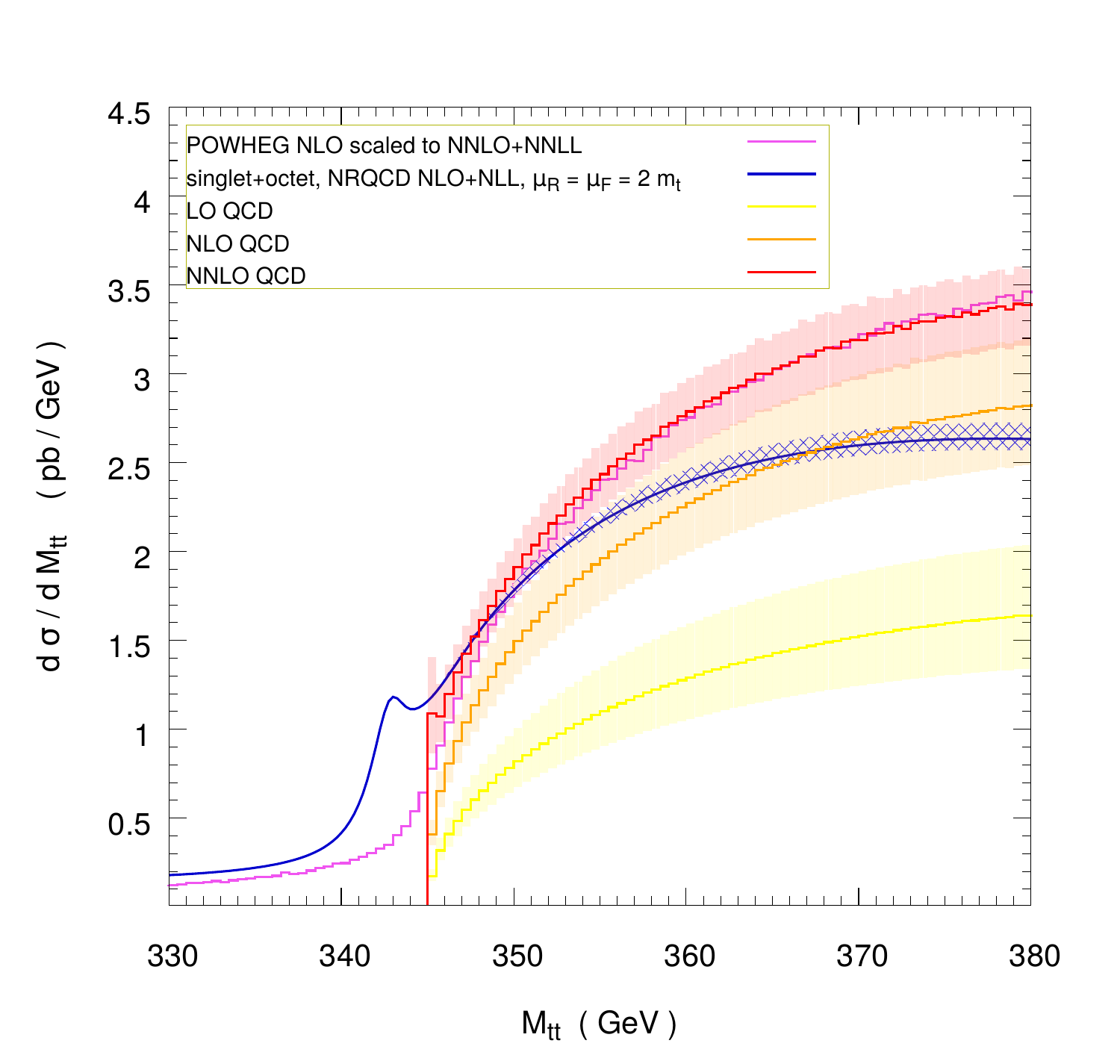}
\caption{
\label{fig:mtt_threshold}
Invariant mass distribution $d\sigma/dM_{t\bar{t}}$ in NRQCD with Coulomb and threshold resummation (blue)
compared to fixed-order perturbative QCD at LO (yellow), NLO (orange) and NNLO (red) starting at a threshold of $2m_t = 345$~GeV.
The results for the \texttt{POWHEG} predictions at NLO with subsequent rescaling produced by the CMS collaboration (purple) are also shown.
}
\end{center}
\end{figure}
\begin{figure}[h]
\begin{center}
\includegraphics[width=0.7\textwidth]{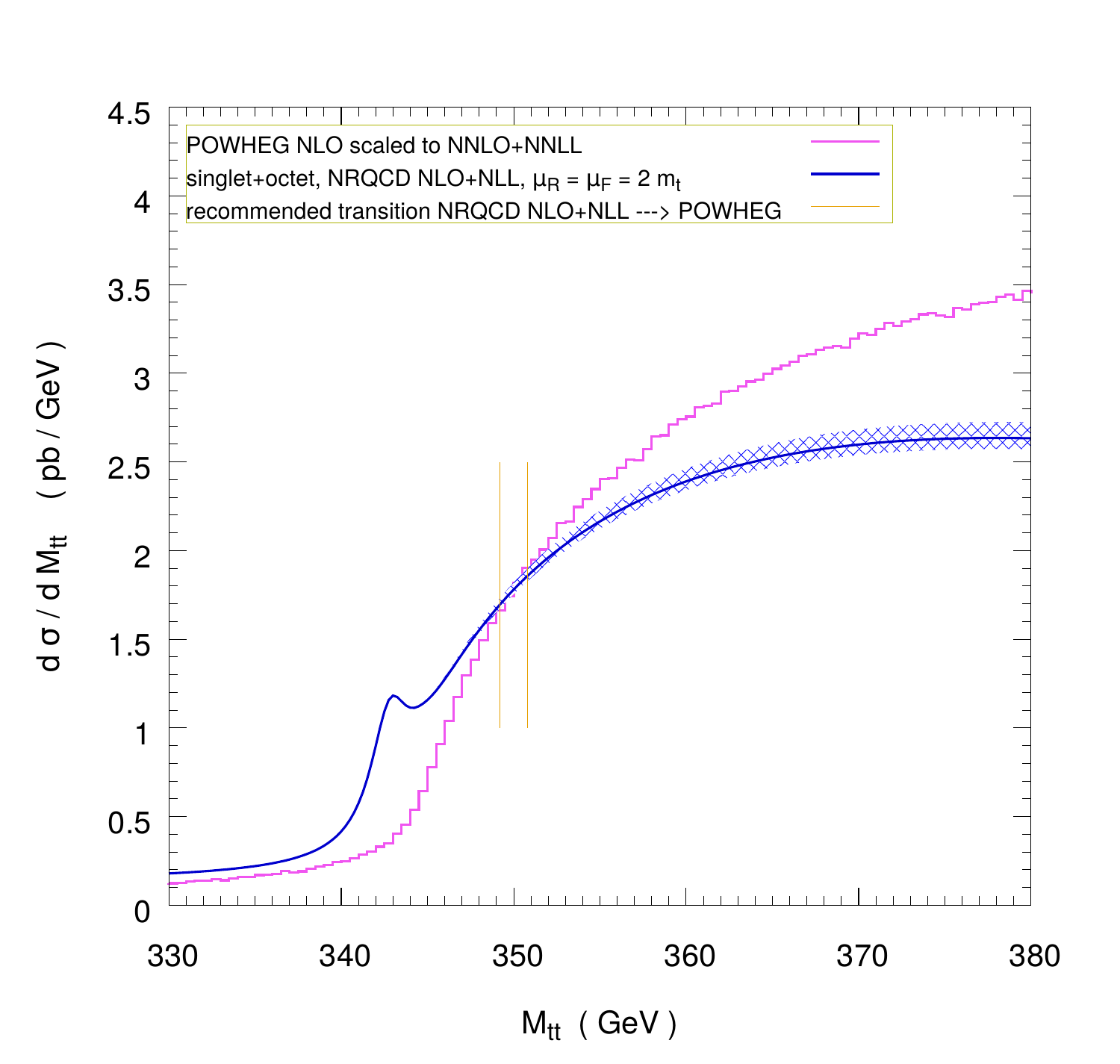}
\caption{
\label{fig:matching}
Same as Fig.~\ref{fig:mtt_threshold} with the matching region for NRQCD predictions (blue) and \texttt{POWHEG} predictions (violet) indicated by vertical lines above  $M_{t\bar{t}} \gtrsim 350$~GeV. 
}
\end{center}
\end{figure}

In order to relate to the recent analyses of the CMS collaboration~\cite{CMS:2024ynj}, 
aimed at searching for pseudoscalar Higgs bosons decaying into $t\bar{t}$ pairs, we compare our NLO+NLL results with their predictions for the SM background from $t\bar{t}$ production in Fig.~\ref{fig:mtt_threshold}.
The theory results for $d\sigma/dM_{t\bar{t}}$ used by CMS rely on predictions obtained with \texttt{POWHEG}~\cite{Nason:2004rx, Frixione:2007vw, Alioli:2010xd}, normalized to the inclusive cross sections from \texttt{Top++}~\cite{Czakon:2011xx} at NNLO and with threshold resummation at next-to-next-to-leading logarithmic (NNLL) accuracy.\footnote{The experimental analysis~\cite{CMS:2024ynj} uses $\Gamma_t = 1.3$ GeV instead of $\Gamma_t = 1.4$ GeV. This has a minor impact on the predictions for the $M_{t\bar{t}}$ distribution and does not change the conclusions of this work. Additionally, the CMS analysis uses a dynamical scale $\mu_R=\mu_F=H_T/2$ instead of the static scale choice applied here. The latter reduces the effect of higher-order contributions, i.e.\ provides better perturbative convergence.
Near threshold $p_T$ is small, so that the numerical value of the dynamical scale $H_T/2 \simeq m_t$, close to the lowest value for fixed scales used in this work.
}  
Subsequently, the results are re-weighted by $K$-factors for the QCD predictions at NNLO from \texttt{MATRIX}~\cite{Catani:2019hip, Grazzini:2017mhc} and the NLO electroweak (EW) corrections from \texttt{Hathor}~\cite{Aliev:2010zk,Kuhn:2013zoa}.
The latter are small and negative, around 2\% only in the kinematic region of interest.
Fig.~\ref{fig:mtt_threshold} shows 
the \texttt{POWHEG} predictions rescaled to NNLO+NNLL accuracy, without NLO EW effects, together with predictions from fixed-order QCD perturbation theory at LO, NLO and NNLO. 
For larger values of $M_{t\bar{t}}$ good agreement with our NNLO results for $d\sigma/dM_{t\bar{t}}$ is observed. 
In the threshold region, our NRQCD predictions for $d\sigma/dM_{t\bar{t}}$ with Coulomb and threshold resummation are clearly enhanced compared to the \texttt{POWHEG} curve.

In Fig.~\ref{fig:matching} we illustrate in an overlay of curves the possible matching between the NRQCD predictions with Coulomb resummation, which are to be applied in range $|M_{t\bar{t}} - 2m_t| \lesssim 5$~GeV, and the \texttt{POWHEG} predictions, applicable at $M_{t\bar{t}} \gg 2m_t$. 
Fig.~\ref{fig:matching} indicates a natural transition region in the range above $M_{t\bar{t}} \gtrsim 350$~GeV, where NRQCD ceases to be valid. 
A more refined treatment of this matching requires dedicated computations beyond the scope of this work. 

For modelling resonances, the CMS collaboration uses a simplified approach, with bound state effects treated according to Ref.~\cite{Fuks:2021xje}.
This model is limited to adding a ${t\bar{t}}$ pair in a ${^1}S_0^{[1]}$ state to the \texttt{POWHEG} samples. 
It is realized by introducing a generic massive spin-0, color-singlet (pseudoscalar) state $\eta_t$ coupled to gluons and top quarks, whose mass and decay width are fitted, and whose impact is restricted to the $M_{t\bar{t}} \in$ [337, 349] GeV region.

The model of Ref.~\cite{Fuks:2021xje} (and its recent refinement~\cite{Fuks:2024yjj}, which uses re-weighting of matrix elements with a NRQCD Green's function) 
can be easily implemented in Monte Carlo codes.
Monte Carlo generators serve complementary purposes, such as predicting general observables and simulating full events.
But in the versions developed until now they omit QCD effects at higher orders, the repulsive interaction among all ${t\bar{t}}$ pairs in color octet states, 
as well as other $S$ (and $P$)-wave singlet states.
As an alternative, for the invariant mass distribution covered in this work,
to improve the description of the ${t\bar{t}}$ bound-state effects, we propose to use our NRQCD predictions with Coulomb and threshold resummation for the dominant channels, also listed in Tab.~\ref{tab:nlo-res}. 
In the range 340~GeV $ \lesssim M_{t\bar{t}} \lesssim 350$~GeV these will provide well-defined theoretical input to the experimental analyses. 
For the time being, we propose a simple additive matching of our NRQCD predictions with perturbation theory at NNLO above $M_{t\bar{t}} \gtrsim 350$~GeV as indicated in Figs.~\ref{fig:mtt_threshold} and \ref{fig:matching}. 
The integrated cross section for the NLO+NLL resummed predictions in the bin $M_{tt} \in [340, 350]$~GeV amounts to 12 pb for the central scale choice $\mur=\muf=2m_t$.
In comparison, the {\tt POWHEG} predictions, integrated in the same range, give 8.4 pb. 
The differences between Ref.~\cite{Fuks:2021xje} and our NRQCD approach can be significant with impact on the shape, width and height of the peak associated with the ${t\bar{t}}$ bound state. 
With solid modelling of the SM background, BSM searches, e.g.\ for a pseudoscalar Higgs boson, can credibly quantify any additional excess in data. 

Within the current accuracy, i.e.\ using the NLO QCD potential for Coulomb resummation and performing threshold resummation at NLL order, our NRQCD predictions are subject to a number of uncertainties.
One dominant issue is that we are missing out on higher-order corrections in perturbation theory in the computation of the Green's function, with the effect of the NNLO corrections estimated to increase the cross section by ${\cal O}(10\%)$ in the peak region.
Another sizeable positive contribution of a similar magnitude, i.e., ${\cal O}(10\%)$ at threshold, is expected from the NNLO corrections to the hard functions $F_{ij \to T}$, as well as from the systematic all-order treatment that addresses their missing NLP threshold corrections in Mellin $N$ space. 
This estimate is commensurate with the scale uncertainties for the hadronic cross section. 
Both these effects are expected to enhance the cross section, so that our current NRQCD prediction can be considered as an estimate for the lower bound on $d\sigma/dM_{t\bar{t}}$ in region $M_{t\bar{t}} \simeq 2m_t$.

The impact of different parameter settings (e.g., scales, values of $m_t$, $\alpha_s(M_z)$, PDFs, etc.) will be covered in an extensive forthcoming study. Theoretical improvements to the approach -- such as accounting for NNLO corrections to the QCD potential and Green's functions, threshold resummation at NNLL order, and a systematic consideration of NLP contributions -- require substantially more work and will be addressed separately in the future.

The NLO+NLL results for $d\sigma/dM_{t\bar{t}}$ are included as an ancillary file with the submission of this work. They are also available from the authors upon request.

\subsection*{Acknowledgements}
We are grateful to C.~Schwanenberger, L.~Jeppe, A.~Grohsjean and A.~Anuar
for discussions on details of the CMS experimental analyses, as well as for providing the \texttt{POWHEG} predictions shown in this work.  
The work of M.V.G.\ and S.M.\ has been supported in part by the Deutsche Forschungsgemeinschaft through the Research Unit FOR 2926, {\it Next  Generation pQCD for Hadron Structure: Preparing for the EIC}, project number 40824754. 
The work of G.L.\ and O.Z.\ has been supported by the Alexander von Humboldt foundation. 
The work of M.S.\ was supported by the Deutsche Forschungsgemeinschaft (DFG, German Research Foundation) under grant 396021762 --- TRR 257 ``Particle Physics Phenomenology after the Higgs Discovery''.


\providecommand{\href}[2]{#2}\begingroup\raggedright\endgroup

\end{document}